\documentclass[prd,10pt,aps,superscriptaddress,nofootinbib,floatfix,amssymb,amsmath]{revtex4}
\usepackage{slashed}
\usepackage{amsmath}

\usepackage{epsfig}

\newcommand{\beqn}{\begin{eqnarray}}
\newcommand{\eeqn}{\end{eqnarray}}

\begin{document}

\title{Landau Levels in graphene in the presence of emergent gravity}

\author{Z.V.Khaidukov}
\affiliation{Institute for Theoretical and Experimental Physics, B. Cheremushkinskaya 25, Moscow, 117259, Russia}

\author{M.A. Zubkov}
\email{zubkov@itep.ru}
\affiliation{Institute for Theoretical and Experimental Physics, B. Cheremushkinskaya 25, Moscow, 117259, Russia}
\affiliation{Moscow Institute of Physics and Technology, 9, Institutskii per., Dolgoprudny, Moscow Region, 141700, Russia}
\affiliation{Far Eastern Federal University,  School of Biomedicine, 690950 Vladivostok, Russia}
\affiliation{National Research Nuclear University MEPhI (Moscow Engineering
Physics Institute), Kashirskoe highway 31, 115409 Moscow, Russia}

\begin{abstract}
We consider graphene in the presence of external magnetic field and elastic deformations that cause emergent magnetic field. The total magnetic field results in the appearance of Landau levels in the spectrum of quasiparticles. In addition, the quasiparticles in graphene experience the emergent gravity. We consider the particular choice of elastic deformation, which gives constant emergent magnetic field and vanishing torsion. Emergent gravity may be considered as perturbation. We demonstrate that the corresponding first order approximation affects the energies of the Landau levels only through the constant renormalization of Fermi velocity. This correction may be detected, for example, through the Hall conductivity. The degeneracy of each Landau level receives correction, which depends essentially on the geometry of the sample. There is the limiting case of the considered elastic deformation, that corresponds to the uniformly stretched graphene. In this case in the presence of the external magnetic field the degeneracies of the Landau levels remain unchanged.
\end{abstract}

\pacs{}

\date{\today}

\maketitle

\section{Introduction}

The two dimensional Weyl semimetal - graphene - has attracted recently a lot of attention \cite{Katsbook} not only because it has unusual electronic properties. The second important reason is that it may represent the arena for the observation in laboratory of various effects specific for the high energy physics. The low energy physics of graphene is described by relativistic effective action for the fermions. The elastic deformations
\cite{Landau,Bilby1956,Kroener1960,Dzyaloshinskii1980,KleinertZaanen2004,Vozmediano2010,Zaanen2010}
in topological media with Fermi points give rise to emergent gravity \cite{Froggatt1991,Volovik2003,Horava2005}. For graphene the emergent gravity corresponding to elastic deformations was derived in \cite{VolovikZubkov2014,VZ2013} (see also \cite{Oliva2013,Oliva2014}).
It accompanies the emergent gauge field, which was discussed in a number of papers \cite{2,3,4,5,6,7,8,9}. In general case for the strained graphene the emergent gravity gives corrections to various quantities, which are typically calculated using the emergent gauge field only. In particular, in \cite{Volovik:2014kja} the scattering of the quasiparticles on the dislocations has been discussed. It has been shown that the Aharonov - Bohm phase originated from the emergent magnetic flux carried by the dilocation receives the correction originated from the emergent gravitational field.

In the present paper we extend the discussion of possible effects caused by emergent vielbein to the consideration of the Landau Levels. Those Landau levels originate from the emergent (or external) magnetic field. The presence of gravity cannot affect the existence of the lowest Landau level (LLL). Symmetry prompts that such a level with zero energy exists also when gravity is taken into account. The degeneracy of the LLL in the presence of gravity is to be described by a certain type of the Atiyah - Patodi - Singer (APS) theorem \cite{APS}. Its simplified form for the case when gravity is absent is discussed, for example, in \cite{Ansourian} - the theorem relates the number of the zero modes of Dirac - Maxwell operator with the total magnetic flux through the surface. When the definition of Dirac operator is extended to include Riemannian gravity the  expression for the number of zero modes for the surface with boundary is expressed through the surface curvature - dependent term, through the total magnetic flux and through the boundary term \cite{APS}, which should depend essentially on the boundary conditions. The boundary term in general case modifies the expression for the number of zero modes. The case, when gravity is not Riemannian was discussed, for example, in \cite{Kimura:2007xa}, where the boundary terms were not considered. We deal with the graphene surface with boundary, while the gravity is not Riemannian. We assume free boundary conditions and impose on the zero modes the condition that they are localized inside the surface. In the considered particular case the direct counting of such zero modes gives expression that contains the term proportional to the product of the total magnetic flux through the surface and the linear size of the sample. Presumably, this term should have its origin at the boundary term of the corresponding version of the APS theorem. However, we do not consider in the present paper the general formulation of the APS theorem for surfaces with boundary and with the non - Riemannian gravity, and restrict our consideration to the particular configuration of the emergent vielbein and emergent magnetic field.

A priori this is not clear are there the corrections to the higher Landau levels and to their degeneracies. We perform the explicit check for the particular elastic deformation, which gives rise to constant emergent magnetic field and vanishing torsion. It appears, that in this particular case in the leading order in the perturbation caused by gravity the correction to the Landau levels results from the constant renormalization of Fermi velocity. In this approximation the degeneracy of the Landau levels receives correction that depends on the particular form of the sample.

\section{From the tight - binding model to the low energy effective field theory}

Let us recall briefly the basic facts about the low energy field theory of graphene.
We start from the the nearest-neighbor tight-binding Hamiltonian of graphene:
$$\hat{H}_{t}=-t\sum_{\vec{A},i,\sigma}{(a^{\dagger}(\vec{A})_{\sigma}b_{\sigma}(\vec{A}+\vec{b}_{i})+H.c)},$$
Here $a^{\dagger}(\vec{A})$ is the creation operator that refers to the sites of the triangular sublattice A. By $\vec{A}$ we denote the position of an atom of the sublattice A, by  $b(\vec{A}+\vec{b}_{i})$ we denote the electron annihilation operator that refers to the sublattice B, and is located at $\vec{B}=\vec{b}_i+\vec{A}$. By $t$  we denote the hopping parameter. In general case the values of $t$ may be different for different points of the honeycomb lattice. However, first we consider the unperturbed case, when the values of $t$ are equal for all lattice sites. The Brillouin zone is a hexagon with the opposite sides identified. There are two different vertices of the hexagon, that are denoted by $K_+$ and $K_-$. Since the quasiparticle energy vanishes at these points, they represent the Fermi points. The low energy expansion around any of these two Fermi points  $K_\pm$  gives the effective low energy  hamiltonian. At $K^-$ it is reduced to the massless Weyl hamiltonian:
\begin{equation}
H_- = v_{f}\int{d^{2}\vec{r}\bar{\Psi}(\vec{r})(i\sigma^3)(i\sigma^{1}\partial_{x}+i\sigma^{2}\partial_{y})\Psi(\vec{r})}\label{H0}
\end{equation}
where $\sigma^{i}$ are Pauli matrices, while $v_{f}=\frac{3t d}{2}$ is Fermi velocity, $t$ is the hopping parameter, $d$ is the distance between the nearest atoms of carbon. $\Psi(\vec{r})$ is the two - component spinor, which may be interpreted as the left - handed spinor after the  constant linear transformation  $\bar{\Psi}\rightarrow \bar{\Psi}(-i\sigma^3)$.
The Hamiltonian at $K_+$ is reduced to
\begin{equation}
H_+ = v_{f}\int{d^{2}\vec{r}\bar{\Psi}(\vec{r})\sigma^2(i\sigma^3)(i\sigma^{1}\partial_{x}+i\sigma^{2}\partial_{y})\sigma^2\Psi(\vec{r})}\label{H01}
\end{equation}
We interpret it as the Hamiltonian for the right - handed spinors, that are related to the original ones by the transformations $\bar{\Psi}\rightarrow \bar{\Psi}(-i\sigma^2\sigma^3)$, and ${\Psi}\rightarrow \sigma^2\bar{\Psi}$
The external gauge field is taken into account via the substitution $i\partial_{i}\to D_i = i\partial_{i}+A_{i}$. Contrary to the external gauge field the emergent gauge field has opposite signs for the opposite Fermi points.

\section{Landau levels in the presence of constant external magnetic field}

\label{magn}

Here we remind the derivation of the Landau levels in graphene and introduce notations to be used further.  The Landau levels appear in the presence of the uniform magnetic field $\bf H$,  which is directed along the $z$ axis ($H^{i}=\epsilon^{ijk}\partial_{j}A_{k}$). We use the gauge  $A_{x}=-\frac{1}{2}Hy, A_{y}=\frac{1}{2}Hx$ (here it is assumed, that $H>0$) and consider the Hamiltonian for the fermions near the $K_-$ Fermi point $$\hat{H}_-(A_k)=iv_f\sigma^{3}\sigma^{k}[i\partial_{k}+A_{k}]$$
Notice, that in the given paper we adopt relativistic notations with the factors $e$ hidden in the definition of electromagnetic field, and with $c = \frac{h}{2\pi}=1$.

In the remaining part of the paper we will use complex coordinates $z=x+iy,\bar{z}=x-iy$. The corresponding derivatives are defined as   $\partial_{z}=\frac{1}{2}(\partial_{x}-i\partial_{y}), \partial_{\bar{z}}=\frac{1}{2}(\partial_{x}+i\partial_{y})$. We have
\begin{equation}\hat{H}_-=  v_{f}  \begin{pmatrix}
	0 & -2\partial_{z}+\frac{1}{2}H\bar{z} \\
	2\partial_{\bar{z}}+\frac{1}{2}Hz & 0 \\
		\end{pmatrix} =  v_{f}\sqrt{2 H}\begin{pmatrix} 0& a^+ \\
		a & 0
		
		\end{pmatrix}, \quad a = \sqrt{\frac{2}{H}}\partial_{\bar{z}}+\sqrt{\frac{H}{8}} {z} \label{H-}\end{equation}
The zero modes of the Hamiltonian satisfy equation  $\hat{H}\Psi=0$ with $\Psi = (\psi,\psi^{\prime})^T$:
\begin{eqnarray}
&&(-\partial_{z}+\frac{1}{4}H\bar{z})\psi^\prime=0\nonumber\\&& (\partial_{\bar{z}}+\frac{1}{4}Hz)\psi=0 \label{eq0}
\end{eqnarray}
The solution of this system is
\begin{eqnarray}
&&\psi^\prime(x,y) = P_2(z,\bar{z}) e^{\frac{1}{4}H\bar{z}z}\nonumber\\&& \psi(x,y) = P_1(z,\bar{z})  e^{-\frac{1}{4}H\bar{z}z} \label{eq01}
\end{eqnarray}
Here $P_{1,2}$ is polynomial. The component $\psi$ is localized in the bulk while $\psi^\prime$ is localized at infinity.
Therefore, we  neglect $\psi^\prime$.

The simplest way to obtain the complete set of the zero  modes is to  introduce  operators  $b,b^{+}$ such that $[b,b^{+}]=1$, while $[a,b^{+}]=[a^+,b]=[a,b]=[a^+,b^+]=0$. We define
 $$b=\sqrt{\frac{2}{H}}\partial_{{z}}+\sqrt{\frac{H}{8}}\bar{z}$$
 The sequence of the states $\psi^{(n)}_{0}$, $n = 0, ...$ that correspond to the LLL starts from the state $\psi^{(0)}_{0}$, which satisfies
 \begin{equation}
 a \psi^{(0)}_{0}= 0, \quad b \psi^{(0)}_{0} = 0
 \end{equation}
We have
$$\psi^{(0)}_{0}= {\rm const}\, e^{-\frac{1}{4}H\bar{z}z}$$
The remaining LLL states are given by
		$$\psi^{(n)}_{0}={\rm const} \, (b^{+})^{n}\psi^{(0)}_{0}$$

In order to obtain the upper Landau levels and the corresponding states we need to solve equation $$H\psi=E\Psi$$
There are two sequences of solutions that start from
$$ \Psi^{(0)}_0 = \begin{pmatrix}
		\psi^{(0)}_0\\
		0 \\
		  \end{pmatrix}$$
and are given by
\begin{equation}
\langle z, \bar{z} | n,k, \lambda\rangle \equiv \Psi^{(k)}_{\lambda,n}=\frac{1}{\sqrt{2}}\begin{pmatrix}
		\psi^{(k)}_n \\
		\lambda \psi^{(k)}_{n-1}\\
		  \end{pmatrix}, \quad \psi^{(0)}_n = \frac{(a^{+})^{n}}{\sqrt{n!}} \psi^{(0)}_0, \quad \langle z, \bar{z} | n,k\rangle \equiv \psi^{(k)}_n = \frac{(b^{+})^{k}}{\sqrt{k!}} \psi^{(0)}_n\label{eigen}
\end{equation}
where  $\lambda=\pm 1$, and the plus and the minus signs correspond to electrons and holes respectively. Energy is given by
$E_{\lambda,n}^{2}=2 v^2_{f} H n$   	
and
$$E_{\lambda,n}=\lambda v_{f} \sqrt{2 n H}$$

Above we considered the Hamiltonian for positive values of $H$. For the negative values the eigenvalues are $\lambda \sqrt{2 n |H|}$ with the eigenstates given by ($\psi^{(k)}_{-1}=0$):
\begin{equation}
\Psi^{(k)}_{\lambda,n}=\frac{1}{\sqrt{2}}
		\begin{pmatrix}
	\lambda \tilde{\psi}^{(k)}_{n-1}\\	\tilde{\psi}^{(k)}_n
				  \end{pmatrix}, \quad \tilde{\psi}^{(0)}_n = \frac{(\tilde{a}^{+})^{n}}{\sqrt{n!}} \psi^{(0)}_0, \quad \tilde{\psi}^{(k)}_n = \frac{(\tilde{b}^{+})^{k}}{\sqrt{k!}} \tilde{\psi}^{(0)}_n\label{eigen2}
\end{equation}
with
\begin{eqnarray}
\tilde{a} &=&- \sqrt{\frac{2}{|H|}}\partial_{{z}}- \sqrt{\frac{|H|}{8}} \bar{z} \nonumber\\
\tilde{b}& = &  \sqrt{\frac{2}{|H|}}\partial_{\bar{z}}+ \sqrt{\frac{|H|}{8}} {z}
\end{eqnarray}

The Hamiltonian $H_+(A_k) = \sigma^2 H_-(A_k) \sigma^2$ for the fermions living near the $K_+$ Fermi point may be considered in the similar way. In the presence of the external gauge field the corresponding spectrum is identical to that of the Hamiltonian at $K_-$.

In the presence of the emergent gauge field, which corresponds to the constant emergent magnetic field $H$ the Hamiltonian at the $K_+$ Fermi point remains the same while the Hamiltonian at $K_-$ differs from Eq. (\ref{H-}) by the change $H \rightarrow - H$.

\section{Influence of elastic deformations on the low energy effective theory of graphene}

In the presence of elastic deformations Eqs. (\ref{H0}) and (\ref{H01}) are modified as follows \cite{VolovikZubkov2014}:
\begin{eqnarray}
H_- &=& \int{d^{2}\vec{r}\, V_f(\vec{r})\, \bar{\Psi}(\vec{r})e^k_a(i\sigma^3)i\sigma^{a}D^{(-)}_{k}\Psi(\vec{r})}\label{H011}
\nonumber\\H_+ &=& \int{d^{2}\vec{r}\,V_f(\vec{r}) \, \bar{\Psi}(\vec{r})e^k_a\sigma^2(i\sigma^3)i\sigma^{a}D^{(+)}_{k}\sigma^2\Psi(\vec{r})}\label{H012}
\end{eqnarray}
Here $$ D^{(\pm)}_k = \partial_k \mp i A_k$$
and $A_k$ is the emergent gauge field, $e_a^k$ is the zweibein matrix with unit determinant, while $V_f$ is the space dependent Fermi velocity.

In the case of weak elastic deformations the hopping parameters differ from each other for different points on the graphene sheet. The simplest model relates the hopping parameters with the tensor of elastic deformations in such a way, that the hopping parameter corresponding to the jump between the two sites ${\bf x}$ and ${\bf x}+{\bf b}_j$ depends only on the real distance between these two sites given by ${r}({\bf x},{\bf b}_j) = |{\bf u}({\bf x}+{\bf b}_j)-{\bf u}({\bf x})|$, where $\bf u$ is the displacement vector. We define the linearized tensor of elastic deformations as follows
\begin{equation}
u_{km} = \frac{1}{2}(\partial_k u_m + \partial_m u_k)
\end{equation}
and introduce the Gruneisen parameter $\beta$ \cite{3}.
The emergent gauge fields are given by
$$A_{y}=\frac{\beta }{2d}(u_{yy}-u_{xx})$$
$$A_{x}=-\frac{\beta }{d}u_{xy}$$
As it was already mentioned, if we consider in the same way the fermions near the second Fermi point, the corresponding emergent gauge field will have the opposite sign.

The emergent zweibein $e^k_a$ and the space dependent Fermi velocity were derived in \cite{VolovikZubkov2014} and for both Fermi points are given by $$e^{i}_{a}=\Bigg(\delta^{i}_{a}(1+\frac{\beta}{2}u_{aa})-\beta \begin{bmatrix} u_{11} & u_{21} \\
u_{12}  & u_{22} \\
 \end{bmatrix}   \Bigg)$$ for $i,a = 1,2$ and
 $$ V_f = v_f (1-\frac{\beta}{2}(u_{11}+u_{22}))$$ Notice, that this expression for the emergent vielbein refers to the so - called accompanying reference frame, in which the coordinates that parametrize the positions of the Carbon atoms are the same as the coordinates in the unperturbed graphene. In the three - dimensional notations the vielbein has the mentioned above zweibein as its special components while $e^0_0 = 1/V_f$, and the determinant of the three - dimentional matrix $e_a^j$ is given by ${\rm det}_{3\times 3}\,e = \frac{1}{V_f}$.

 In the following we will consider the particular configuration of elastic deformations that gives constant value of emergent magnetic field $H$. This configuration is derived in Appendix A and is given by
 \begin{equation}
 u_x=\frac{Hd}{2\beta}(cx+y^2/2-x^2/2), \quad u_{y}=\frac{Hd}{2\beta}(cy+xy) \label{uxuy}
 \end{equation}
 with arbitrary constant $c$.
Notice, that this theory is applicable for the small deformations only. The deformations are small if $$dHc , dHR<<1$$
where $R$ is the linear size of the sample. The important property of the configuration of Eq. (\ref{uxuy}) is that torsion $T^a_{ij} = \partial_i e^a_j - \partial_j e^a_i$ vanishes. In order to demonstrate this let us calculate the vielbein $e^i_a$ and its inverse $e^a_i$:
\begin{equation}
e^{i}_{a}=\Bigg(\delta^{i}_{a}-\frac{Hd}{2}\begin{bmatrix}-x & y   \\
 y  & x\\
 \end{bmatrix}   \Bigg)\label{e}
\end{equation}
 and
 $$e^{a}_{i}\approx \Bigg(\delta^{i}_{a}+\frac{Hd}{2}\begin{bmatrix}-x & y   \\
 y  & x\\
 \end{bmatrix}   \Bigg)$$
The direct calculations using the last expression give $T^a_{ij} = 0$ in the approximation linear in the tensor of elastic deformations.

If we would neglect the nontrivial part of the vielbein and set $e^i_a = \delta^i_a$, then the dynamics of the quasiparticles is described by expressions of Sect. \ref{magn}. Then we should take into account nontrivial contribution to the vielbein as perturbation. In this situation the spectrum of the quasiparticles may be modified. Instead of the given Landau level $E_{\lambda,n}$ with huge degeneracy the several energy levels or several pieces  of continuous energy spectrum may appear. A priori this is not clear what will happen to the Landau levels in the presence of gravity even if torsion vanishes. In the following sections we will demonstrate, that at least in the leading order in tensor of elastic deformations the energies of the Landau levels remain unchanged while their degeneracies receive corrections.

\section{Influence of the emergent gravity on the lowest Landau level}

In the following we will discuss the consequences for the Landau level spectrum that follow from the emergent gravity given by the nontrivial emergent vielbein $e^k_a$. We take for the reference the particular elastic deformations given by Eq. (\ref{uxuy}). In the present section the lowest Landau level (LLL) will be considered.

The elastic deformations of Eq. (\ref{uxuy}) give rise to constant emergent magnetic field $H$ and to vanishing torsion. However, the vielbein remains nontrivial and is given by Eq. (\ref{e}).
The Hamitonian at $K_+$ takes the form
$$\sigma^2H_+\sigma^2=i\sigma^{3}V_f\sigma^{a}e^{k}_{a}\circ[i\partial_{k}+ A_{k}]=i\sigma^{3}V_f\sigma^{a}[e^{k}_{a}i\partial_{k}+\frac{i}{2}(\partial_{k}e^{k}_{a}) + e^{k}_{a}A_{k}]$$

Let us substitute into the Hamiltonian the above expression for the vielbein.
Using complex variables we rewrite the Hamiltonian as follows (see Appendix B)
\begin{equation}
\sigma^2H_+\sigma^2=V_f \begin{pmatrix} 0 & -2\partial_{z}+\frac{H}{2}\bar{z}-z H d\partial_{\bar{z}}-\frac{1}{4}HdHz^{2} \\
2\partial_{\bar{z}}+\frac{H}{2}z+\bar{z}{H}d\partial_{z}-\frac{1}{4}HdH\bar{z}^{2} & 0 \\
\end{pmatrix}\label{HA}
\end{equation}

The zero modes satisfy equations:
$$ (2\partial_{\bar{z}}+\frac{H}{2}z+\bar{z}H d \partial_{z}-\frac{1}{4}HdH\bar{z}^{2})\psi^\prime=0 $$ and $$ (-2\partial_z+\frac{H}{2}\bar{z}-z H d\partial_{\bar{z}}-\frac{1}{4}HdH{z}^{2})\psi=0 $$
As above, the nonzero solution of the second equation is localized at infinity.
The first equation has the particular solution
$$\psi_0^{(0)} = exp\Big(-\frac{H}{4}z\bar{z}+\frac{H^2d}{12}\bar{z}^{3}\Big)$$
 It appears, that operator $\hat{b}$ annihilates $\psi_0^{(0)}$ as well.
One can easily find that the general solution of equation $\hat{A}\psi = 0$ has the form:
\begin{equation} \psi_f = f\Big(z - \frac{Hd}{4}\bar{z}^2\Big)\psi_0^{(0)} \label{pf}\end{equation}
where arbitrary function $f$ satisfies
$2\partial_{\bar{z}}f+ \frac{zHd}{2}\partial_{z}f=0$ simply because of the choice of the argument.

The Hamiltonian at $K_-$ has the form
$$H_-=V_f  \begin{pmatrix} 0 & -2\partial_{z}-\frac{H}{2}\bar{z}-z H d\partial_{\bar{z}} +\frac{1}{4}HdHz^{2}  \\
2\partial_{\bar{z}}-\frac{H}{2}z+\bar{z}{H}d\partial_{z} +\frac{1}{4}HdH\bar{z}^{2} & 0 \\
\end{pmatrix}
$$

The zero modes satisfy equations:
$$ (2\partial_{\bar{z}}-\frac{H}{2}z+\bar{z}H d \partial_{z}+\frac{1}{4}HdH\bar{z}^{2})\psi=0 $$ and $$ (-2\partial_z-\frac{H}{2}\bar{z}-z H d\partial_{\bar{z}}+\frac{1}{4}HdHz^{2} )\psi^\prime=0 $$
Now the nonzero solution of the first equation is localized at infinity.
The second equation has the particular solution
$$\tilde{\psi}_0^{(0)} = exp\Big(-\frac{H}{4}z\bar{z}+\frac{H^2d}{12}{z}^{3}\Big)$$
The general solution for the zero mode has the form:
\begin{equation} \tilde{\psi}_f = f\Big(\bar{z} - \frac{Hd}{4}{z}^2\Big)\tilde{\psi}_0^{(0)}\label{pf2} \end{equation}
where function $f$ is arbitrary.

\section{Degeneracy of the lowest Landau level}

Arbitrary functions $f$ enter the expressions for the wave functions of the zero modes Eqs. (\ref{pf}), (\ref{pf2}). In order to construct the complete set of the wave functions we may start from
$f_n(z) = z^n$ for $n = 0, 1, ...$, and orthogonalize this set with the weight $|\psi_0^{(0)}|^2$ at $K_-$ and correspondingly, with $|\tilde{\psi}_0^{(0)}|^2$ at $K_+$. For example, $f_1 \rightarrow f_1 - f_0 \frac{( f_1|f_0 )}{( f_0|f_0)}$. The highest degree of $z$ in $f_n$ after the application of this procedure remains equal to $n$. Say, at $K_+$ the values of the scalar products are given by $( f_i|f_j ) = \int d \bar{z} dz \bar{f}_i(z)f_j(z)e^{-\frac{H}{2}\bar{z}z +{\rm Re}\,\frac{H^2d}{6}{z}^{3}}$. They depend essentially on the boundary conditions. However, in order to count the number of the zero modes we do not need to know the details of these conditions.

Let us suppose, that $H>0$, while graphene surface has the form of the circle with radius $R$. The boundary conditions at $r=R$ are free. Then the number of the zero modes $N$ for the fermions at $K_+$ may be estimated as the number of the function $f_N$ such that the corresponding wave function has its maximum at $|z| = R$. In radial coordinates $z = r e^{i\phi}$ we have
$$ |\psi_{f_N}(r,\phi)|^2 \sim r^{2N}(1-\frac{Hd}{4} r e^{3i\phi})^{N}(1-\frac{Hd}{4}re^{-3i\phi})^{N}exp\Big(-\frac{H}{2}r^{2} +\frac{H^2d}{12}r^{3}(e^{i3\phi}+e^{-3i\phi})\Big)$$
We need to find the position of the maximum of expression
\begin{eqnarray}
{\rm log} \,|\psi_{f_N}(r,\phi)|^2 &\sim & 2N {\rm log}(r)+N {\rm log}(1-\frac{Hd}{2}r\cos(3\phi))-\frac{1}{2}Hr^{2}+\frac{H^2d}{6}r^{3}\cos(3\phi)\nonumber\\ &\sim & 2N {\rm log}(r)-\frac{1}{2}Hr^{2}+\Big(\frac{H^2d}{6}r^{3}-N \frac{Hd}{2}r\Big)\cos(3\phi)
\end{eqnarray}
The zero order approximation (which corresponds to the case when the gravitational contribution is neglected) gives
$$N^{(0)} = H R^2/2 = \frac{H \pi R^2}{2 \pi}$$
that is the number of the zero modes is equal to the normalized magnetic flux through the sample.

When the gravitational corrections are taken into account maximum of the wave function is achieved when $\cos (3 \phi) = -1$. This gives
\begin{equation}
N\Big(1+\frac{HdR}{4}\Big)=\frac{1}{2}HR^{2}\Big(1+\frac{HdR}{2}\Big) \label{09}
\end{equation}
Thus  in the leading approximation  we arrive at the following  expression: $N = \frac{\int dx d y H}{2\pi}\Big(1+\frac{HdR}{4}\Big)$.
The direct calculation gives the same result for the fermions at the Fermi point $K_+$.

Recall, that above we have calculated the degeneracy for the positive values of $H$. For negative $H$ the calculation is similar and it gives
\begin{equation}
N = \frac{\int dx d y |H|}{2\pi}\Big(1+\frac{|H|dR}{4}\Big)\label{NH}
\end{equation}

It is worth mentioning, that if we turn on the external magnetic field $B$ in addition to the emergent magnetic field $H$ the result for the degeneracy of the LLL at $K_\pm$ is given by $N = \frac{\int dx d y |\pm H+B|}{2\pi}\Big(1+\frac{|H| dR}{4}\Big)$.

Thus the degeneracy of the LLL receives correction that is proportional both to the interatomic distance $d$ and the linear size of the sample $R$. This is an interplay between the infrared and the ultraviolet cutoffs of the low energy effective theory. The dependence on the infrared cutoff $R$ means, that the particular form of the sample influences essentially the correction to the degeneracy.

\section{Corrections to the higher Landau levels}

The wave functions of the Landau levels of arbitrary order near to the Fermi point $K_+$ in the absence of gravity are given by Eq. (\ref{eigen}). The corrections due to gravity result from the renormalization of Fermi velocity $v_f \rightarrow V_f=v_f(1-\frac{Hdc}{2})$ (where constant $c$ enters the expression for the elastic deformations) and also from the  perturbation
$$\sigma^2\hat{V}_+\sigma^2=-V_f H d \begin{pmatrix} 0 & (a+b^{+})a \\
  (a^{+}+b)a^{+} & 0\\

\end{pmatrix}
$$
Its matrix elements that determine the first order corrections to the $n$ - th level are proportional to
$$<n-1,m|(a^{+}+b)a^{+}|n,q>+h.c=\sqrt{n+1}<n-1,m|a^{+}+b|n+1,q>+h.c=$$
$$=(\sqrt{(n+1)(n+2)}<n-1,m|n+2,q>+\sqrt{(n+1)q}<n-1,m|n+1,q+1>)+h.c$$
From this expression one can easily see, that the Landau Levels have no first order corrections (except the corrections hidden in the Fermi velocity $V_f$).

Let us define
$$ \hat{A} = a -\sqrt{\frac{H}{2}}d([\hat{a}^+]^2+\hat{b} \hat{a}^+)$$
One can see, that equation for the zero mode $\psi_0^{(0)}$ is written as $\hat{A}\psi_0^{(0)} = 0$ while $$[\hat{A},\hat{A}^+]=1 + H d^2 \frac{z\partial_z - \bar{z}\partial_{\bar{z}}- H \bar{z}z}{2}$$
Let us introduce the operator $B^+$, that produces the sequence of the zero modes $\psi_0^{(k)}$ described in the previous section starting from $\psi_0^{(0)}$. (We do not need its explicit form here.) Now the eigenvectors of Hamiltonian corresponding to the eigenvalues $$E_{\lambda,n} = \lambda V_f \sqrt{2n H}=\lambda v_f(1-\frac{1}{2}Hdc)\sqrt{2n H}, \quad \lambda = \pm 1$$ may be constructed as follows (up to the terms linear in the lattice spacing $d$)
\begin{equation} \Psi^{(k)}_{\lambda,n}=\frac{1}{\sqrt{2}}\begin{pmatrix}
		\psi^{(k)}_n \\
		\lambda \psi^{(k)}_{n-1}\\
		  \end{pmatrix}, \quad \psi^{(k)}_n = \frac{(\hat{A}^{+})^{n}}{\sqrt{n!}} \psi^{(k)}_0, \quad \psi^{(k)}_0 = \frac{(B^{+})^{k}}{\sqrt{k!}} \psi^{(0)}_0\label{eigen2}
\end{equation}
with $\psi_0^{(0)} = exp\Big(-\frac{H}{4}z\bar{z}+\frac{H^2d}{12}\bar{z}^{3}\Big)$. The second term in the commutator $[\hat{A},\hat{A}^+]$ gives corrections  proportional to $d^2$ both to the energy levels and to the eigenfunctions of the Hamiltonian. We neglect these corrections in our study.
In the case, when the size of the sample is sufficiently large, for the observed Landau levels  $n \ll k$, and we may estimate the degeneracy of each Landau level in the same way as we did for the LLL. This gives the same expression Eq. (\ref{NH}) in the leading order.

In the similar way the same conclusion is reached for the fermions living at $K_-$. The inclusion into consideration of the constant external magnetic field $B$ is straightforward. In this case we have the Landau levels at $K_\pm$ with energies $\lambda v_f\sqrt{2 n |B\pm H|}$ with the degeneracy $\frac{\int d^x d y |B \pm H|}{2\pi}\Big(1+\frac{|H|dR}{4} \Big)$.

Let us also notice the conditions, under which our consideration of the possible corrections to Landau levels is relevant:
\begin{equation}
1 \ll |B \pm H| R^2, \quad |H|dR \ll 1, \quad |H|dc \ll 1
\end{equation}

\section{Possible experimental detection of the considered gravitational corrections}

There is the important simple limiting case of the configuration considered in the present paper. In this limiting case the emergent magnetic field vanishes, while the product $\gamma = cHd$ remains nonzero. This is the case of the uniformly stretched graphene (with the scaling coefficient $\frac{\gamma}{2\beta}$). In this case the degeneracy of the Landau levels (caused by external magnetic field $B$) receives no corrections while the energies of the Landau levels become
\begin{equation}
E_{\pm,n} = \pm v_f(1-\frac{\gamma}{2})\sqrt{2n |B|}\label{E1}\end{equation}
This suggests the simple quantum Hall experiment. In the presence of external electric field the Hall conductivity is given by
\cite{Katsbook}:
\begin{equation}
\sigma_{xy} = \frac{2}{\pi}\Big(\frac{1}{2}+ \sum_{n\ne 0}\theta(|\mu| - E_{+,n})\Big)\, {\rm sign}(\mu)\label{sigma}
\end{equation}
(In order to obtain conductivity in the usual system of units we should multiply this expression by $2\pi e^2/h$.) Here $\theta(x) = \frac{1}{2}(1+{\rm sign}(x))$ is the step function.
For the given value of chemical potential $\mu$ this expression gives the dependence of Hall conductivity on external magnetic field that differs from the conventional one due to the factor $1-\gamma/2$ in Eq. (\ref{E1}). 

When the emergent magnetic field does not vanish, the expected expression for the Hall conductivity becomes more complicated. We have the two sequences of the Landau levels (for each of the two Fermi points $K_\pm$):
\begin{eqnarray}
E^{(K_+)}_{\pm,n} &=& \pm v_f(1-\frac{\gamma}{2})\sqrt{2n |B + H|}\nonumber\\ 
E^{(K_-)}_{\pm,n} &=& \pm v_f(1-\frac{\gamma}{2})\sqrt{2n |B - H|}\nonumber
\end{eqnarray}
and the factors $(1+{\cal G})$ that change the degeneracy of the levels. For the idealized graphene sample of the form of the circle with free boundary conditions the factors ${\cal G}$ were calculated above and are given by ${\cal G} = \frac{1}{4}|H|dR$. For the real graphene sample those factors are different but we expect that they depend on the sample size linearly. We come to the following expression for the Hall conductivity in this case:
\begin{equation}
\sigma_{xy} = \frac{1}{\pi}\Big(1+ \sum_{n\ne 0}\theta(|\mu| - E^{(K_+)}_{+,n})+ \sum_{n\ne 0}\theta(|\mu| - E^{(K_-)}_{+,n})\Big)\,(1+{\cal G}) \, {\rm sign}(\mu)\label{sigma2}
\end{equation}
Here the sequence of the plateaus is doubled compared to the case, when there is no emergent magnetic field due to the different values of the Landau level energies for the opposite Fermi points.

Notice, that the same perturbed values of the Landau level energies may be observed using the investigation of the periodical dependence of various macroscopic quantities on magnetic field - for example through the observation of peaks in longitudinal conductivity (Shubnikov - de Haas effect).

\section{Conclusions and discussion}

To conclude, we considered graphene in the presence of the elastic deformation of particular form, which gives constant emergent magnetic field. This elastic deformation also causes the appearance of emergent gravity with vanishing torsion. In the absence of gravity the well - known (relativistic) Landau levels appear in the spectrum of the fermionic quasiparticles. A priori this was not clear what happens to those Landau levels when gravity is taken into account. In principle, one might expect, that instead of each single Landau level with huge degeneracy the set of discrete energy levels (or the set of the pieces of continuum energy spectrum) may appear.

We investigate by direct calculation of spectrum the influence of the emergent gravity on Landau levels and on their degeneracy. It appears, that in the same approximation, in which the expression for the emergent vielbein was derived in \cite{VolovikZubkov2014}  the energies of Landau levels are affected by gravity only through the constant renormalizaion of Fermi velocity. This conclusion is also not altered if the constant external magnetic field is taken into account. The degeneracy of the Landau levels receives correction, which is proportional to the interatomic distance, to the linear size of the sample, and to the emergent magnetic field. The dependence on the linear size of the sample means that the precise expression of this correction depends essentially on the particular form of the graphene sample.

There exists the important limiting case of the configuration considered in the present paper, when the emergent magnetic field vanishes, while the product $\gamma = cHd$ remains nonzero. This is the case of the uniformly stretched graphene. In this case the degeneracy of the Landau levels (caused by the external magnetic field) receives no corrections while the energies of the Landau levels acquire corrections due to gravity. This suggests the detection of the given corrections through the quantum Hall effect. Magnitude of the Hall conductivity should remain unchanged for the stretched graphene while the dependence on the magnetic field should manifest the plateaus at the values of magnetic field different from the ones observed for the unperturbed graphene. Those corrected values of Landau level energies may be detected experimentally also through the consideration of the other quantities, - for example, through the Shubnikov - de Haas effect. 

In the situation, when both the external and the emergent magnetic fields are present we expect the two complications of the quantum Hall effect based on our study of the particular elastic deformation. First, the Landau levels are doubled due to the presence of the contributions to the total magnetic field due to the emergent magnetic field of opposite signs for opposite Fermi points. This will shift the positions of the plateaus of the Hall conductivity (and, say, the positions of peaks of the Shubnikov - de Haas oscillations).  Besides, our Eq. (\ref{sigma2}) indicates, that the value of Hall conductivity will be enhanced by the factor that accounts for the change of the Landau level degeneracy. It is worth mentioning, that we did not study the elastic deformations of arbitrary form. For the case, when the emergent magnetic field is present we restricted ourselves by the particular configuration of elastic deformation given by Eq. (\ref{uxuy}). Therefore, for the elastic deformation of the general form we may expect the further complication of the pattern of quantum Hall effect and the Shubnikov - de Haas effect as well as the more complicated pattern of the dependence on external magnetic field of the other macroscopic quantities.

The authors kindly acknowledge useful comments by Maurice Oliva-Leyva. It is worth mentioning, that the present work has appeared as a response to the advise by Patrick Lee addressed to one of the authors (M.A.Z.) to investigate the influence of emergent gravity on the Landau levels in graphene. The  work of M.A.Z. was partially supported by  Ministry of science and education of Russian Federation under the contract 02.A03.21.0003,  and by grant RFBR 14-02-01261. The work of Z.V.K. was supported by grant RFBR 14-02-01185.

\section*{Appendix A. Elastic deformations that correspond to the constant emergent magnetic field}

Let us find the configuration that corresponds to the constant magnetic field. We will not try to  find all possible configurations of the strain field, which corresponds to the constant magnetic field and will obtain the particular configuration to be used later for the investigation of the influence of emergent gravity on the Landau levels.

The displacement field $u^i_{x}$ we are looking for satisfies the following equation
$$\partial_{x}A_{y}-\partial_{y}A_{x}=\frac{\beta }{2a}(\partial_{x}(u_{yy}-u_{xx})+2\partial_{y}u_{xy})=H$$
Let us consider the gauge field of the form
\begin{equation}
A_{y}=(1-l)Hx+\partial_y \alpha(x,y),A_{x}=-lHy+\partial_x \alpha(x,y) \label{eqx}
\end{equation} where $l$ is the real number, while $\alpha(x,y)$ is the real - valued function.
We also assume $u_{x}=\partial_{x}\phi, u_{y}=\partial_{y}\phi$,
which leads to the following equations
$$\frac{\beta}{2d}(\partial^{2}_{y}-\partial^{2}_{x})\phi=(1-l)H x+\partial_y \alpha,$$
$$-\frac{\beta}{d} \partial_{x}\partial_y\phi=-l H y+\partial_x \alpha$$

First, let  us require $$\partial^2_{x}\phi+\partial^2_{y}\phi=0$$
It appears, that even with these constraints the choice of the displacement vector is not unique.
We begin with  the solution for the case $\alpha = 0$.
Then the system of equations can be written as
$$\frac{\beta}{d}\partial^{2}_{x}\phi=-(1-l) H x $$
\begin{equation}
\frac{\beta}{d}\partial^{2}_{y}\phi=(1-l) H x  \label{eq3}
\end{equation}
$$-\frac{\beta}{d}\partial_{x}\partial_y \phi=-l H y $$
From this system of equations we derive
$$\phi=\frac{d}{\beta}(\frac{1}{4} H y^2 x-\frac{1}{12} H x^{3}), \quad l = 1/2$$
and
$$u_x=\frac{d}{\beta}(\frac{1}{4} H y^2 -\frac{1}{4} H x^{2}), \quad u_y  = \frac{d}{\beta} \frac{1}{2} H y x$$
Notice, that this theory is applicable for the small deformations only. The deformations are small if $$dHR<<1$$ where $R$ is the linear size of the graphene sample.

Now let us weaken the above constraint and assume that
$$u_{xx}+u_{yy}=\frac{Hd}{\beta}c,c\ne 0$$ for a certain constant $c$.
Again we set $l=1/2$
$$u_{i}=\partial_{i}\phi,i=x,y$$
In this  case
$$\phi=\frac{Hd}{4\beta}(cx^2+cy^2+xy^2-x^3/3)$$
Then
$$u_{x}=\frac{Hd}{2\beta}(cx+y^2/2-x^2/2)$$
$$u_{y}=\frac{Hd}{2\beta}(cy+xy)$$
For the gauge potentials we have
$$A_{x}=-\frac{\beta}{d}u_{xy}=-\frac{H}{2}y,A_y=\frac{H}{2}x$$

The Fermi velocity is renormalized and has the form
$$V_f=v_f(1-\frac{\beta}{2} (u_{xx}+u_{yy}))=v_f(1-\gamma/2),\quad \gamma = Hdc$$
The zweibein is given by
$$e^{i}_{a}=\Big(1+\frac{\beta}{2}u_{aa}-\beta\begin{pmatrix} u_{xx} & u_{xy} \\
   u_{yx} & u_{yy} \\
																													\end{pmatrix}
 \Big)$$
This gives
\begin{equation}
e^{i}_{a}=\Bigg(\delta^{i}_{a}-\frac{Hd}{2}\begin{bmatrix}-x & y   \\
 y  & x\\
 \end{bmatrix}   \Bigg)\\
\end{equation}

\section*{Appendix B. Correction to the Hamiltonian due to gravity}

Let us start from the calculation of the product
$$e^{k}_{a}A_{k}=\Bigg(\delta^{i}_{a}-\frac{Hd}{2}\begin{bmatrix}-x & y   \\
 y  & x\\
 \end{bmatrix}   \Bigg)\times \begin{bmatrix} -\frac{1}{2}Hy \\
 \frac{1}{2}Hx \\
\end{bmatrix}=\begin{bmatrix} -\frac{1}{2}Hy-\frac{Hd}{2} Hyx \\
\frac{1}{2}Hx-\frac{Hd}{4}(Hx^{2}-Hy^{2})
\end{bmatrix}
 $$
Here we assume $H>0$. The Hamiltonian contains the products of the sigma matrices
$$i\sigma^{3}\sigma^{x}=-\sigma^{y}; i\sigma^{3}\sigma^{y}=\sigma^{x}$$
We have
$$-\sigma^{y}e_{x}^{k}A_{k}=\begin{bmatrix} 0 & i(-\frac{1}{2}Hy-\frac{Hd}{2} Hyx)\\
-i(-\frac{1}{2}Hy-\frac{Hd}{2} Hyx) & 0 \\ \end{bmatrix}
$$
and
$$\sigma^{x}e_{y}^{k}A_{k}=\begin{bmatrix} 0 & \frac{1}{2}Hx-\frac{Hd}{4}(Hx^{2}-Hy^{2}) \\
\frac{1}{2}Hx-\frac{Hd}{4}(Hx^{2}-Hy^{2}) & 0 \\

\end{bmatrix}
 $$
Therefore,
$$-\sigma^{y}e_{x}^{k}A_{k}+\sigma^{x}e_{y}^{k}A_{k}=$$ $$=\begin{bmatrix} 0 & \frac{1}{2}Hx-\frac{Hd}{4}(Hx^{2}-Hy^{2})+i(-\frac{1}{2}Hy-\frac{Hd}{2} Hyx) \\
\frac{1}{2}Hx-\frac{Hd}{4}(Hx^{2}-Hy^{2})-i(-\frac{1}{2}Hy-\frac{Hd}{2} Hyx) & 0 \\
\end{bmatrix}$$
$$=\begin{bmatrix} 0 & \frac{1}{2}H\bar{z}-\frac{1}{4}HdHz^{2} \\
\frac{1}{2}Hz-\frac{1}{4}HdH\bar{z}^{2} & 0 \\
 \end{bmatrix}$$ and the complete Hamiltonian at $K_+$ receives the form
\begin{equation}
\sigma^2\hat{H}_+\sigma^2=V_f \begin{pmatrix} 0 & -2\partial_{z}+\frac{H}{2}\bar{z}-z H d\partial_{\bar{z}}-\frac{1}{4}HdH{z}^{2} \\
2\partial_{\bar{z}}+\frac{H}{2}z+\bar{z}{H}d\partial_{z}-\frac{1}{4}HdH\bar{z}^{2} & 0 \\
\end{pmatrix}=
\end{equation}
\begin{equation}
=V_f \begin{pmatrix} 0 & -2\partial_{z}+\frac{H}{2}\bar{z}-z H d(\partial_{\bar{z}}+\frac{1}{4}Hz)\\
2\partial_{\bar{z}}+\frac{H}{2}z+\bar{z}{H}d(\partial_{z}-\frac{1}{4}H\bar{z}) & 0 \\
\end{pmatrix}
\end{equation}
Recall that $a^{+}=\sqrt{\frac{2}{H}}(-\partial_{z}+\frac{H}{4}\bar{z})$
Therefore,
$$\sigma^2\hat{H}_+\sigma^2=V_f \sqrt{2H}\begin{pmatrix} 0 & a^{+}-zHda/2 \\
  a-\bar{z}Hda^{+}/2 & 0\\

\end{pmatrix}
$$
Using
$$\bar{z}=\sqrt{\frac{2}{H}}(a^{+}+b)$$
we come finally to
$$\sigma^2\hat{H}_+\sigma^2=V_f\sqrt{2H}\begin{pmatrix} 0 & a^{+}-\sqrt{\frac{H}{2}}(a+b^{+})da \\
  a-\sqrt{\frac{H}{2}}(a^{+}+b)da^{+} & 0\\

\end{pmatrix}$$

\end{document}